\documentclass[3p,times,twocolumn]{elsarticle}

\usepackage{ecrc}

\volume{00}

\firstpage{1}

\journalname{Nuclear Physics B Proceedings Supplement}

\runauth{A. Datta et al.}

\jid{nuphbp}

\jnltitlelogo{Nuclear Physics B Proceedings Supplement}

\usepackage{amssymb}

\usepackage[figuresright]{rotating}

\begin{document}

\begin{frontmatter}



\dochead{}

\title{Universal Extra-Dimensional models with
boundary terms: Probing at the LHC}



\author[ad1]{\sf Anindya Datta}
\address[ad1]{Department of Physics, University of Calcutta,  
92 Acharya Prafulla Chandra Road, Kolkata 700009, India}
\author[ad1,ad2]{\sf Ujjal Kumar Dey}  
\address[ad2]{Harish-Chandra Research Institute, 
Chhatnag Road, Jhunsi, Allahabad  211019, India}
\author[ad1,ad3]{\sf Amitava Raychaudhuri}
\address[ad3]{(Conference speaker)}
\author[ad1]{\sf Avirup Shaw}


\begin{abstract}

In universal extra-dimensional models a conserved $Z_2$ parity
stabilizes the lightest Kaluza-Klein particle, a
dark-matter candidate.  Boundary-localized kinetic
terms, in general, do not preserve this symmetry. We examine,  in
the presence of such terms, the single production of Kaluza-Klein
excitations of the neutral electroweak gauge bosons and their
decay to zero-mode fermion-antifermion pairs. We explore how
experiments at the Large Hadron Collider constrain
the boundary-localized kinetic terms for different 
compactification radii.

\end{abstract}

\begin{keyword}


Extra dimension \sep Boundary-localized terms

\end{keyword}

\end{frontmatter}



\section{Introduction} 
\label{intro}

The Large Hadron Collider (LHC) at CERN is a valuable window
for exploring particle physics models.  Here we look at a class of
models where {\em all} particles have access to an
extra dimension \cite{acd}.  The models which we examine 
can be termed `non-minimal universal extra-dimensions' (nmUED)
for reasons discussed below. We show 
that  LHC  running at 8 TeV may exclude significant
segments of the parameter space of this class of models. 

We consider one extra spacelike dimension, $y$, which is flat and
compactified. This coordinate runs from 0 to $2 \pi R$, where $R$
is the radius of compactification. All particles -- scalars,
spin-1/2 fermions, and gauge bosons -- are  five-dimensional
fields or equivalently towers of four-dimensional Kaluza-Klein
(KK) states. The $n$-th level KK states for all particles have
the same mass $n/R$. Further, to retain fermion chirality a
${Z}_2$ symmetry ($y \leftrightarrow -y$) is imposed.  Thus the
extra dimension is compactified on an orbifold $S^1/Z_2$ with  $y
\equiv y +  \pi R$.   This symmetry results in a conserved
KK-parity given by $(-1)^n$ where $n$ is the KK-level. The
standard model (SM) particles have $n$ = 0 and are of
even parity while the  KK-states of the first level are odd. 
KK-parity ensures that the lightest $n = 1$
particle is absolutely stable and hence a  dark matter
candidate, the Lightest Kaluza-Klein Particle (LKP). This
constitutes  the Universal Extra Dimension (UED)
Model.

$S^1/Z_2$ has two fixed points at $y = 0$ and $y = \pi R$. One
can admit additional interaction terms between the KK-states at
these points. These are commonly introduced as counterterms
for 5-dimensional loop-induced effects. In the minimal Universal
Extra-Dimensional Models (mUED) \cite{cms1} these terms are so
chosen that 5-dimensional loop contributions at the cutoff scale
$\Lambda$ are exactly compensated and the corrections, e.g.,
logarithmic contributions to masses of KK particles, can be taken
to be zero at $\Lambda$. These  contributions remove the mass
degeneracy among states at the same KK-level $n$.

In this work \cite{ddrs1} we examine non-minimal UED (nmUED).
Here the boundary terms are not restricted to the special choice
in mUED.  The two departures are:  the two boundary terms are
allowed to be unequal -- this breaks a remnant  $Z_2$ symmetry
which exchanges $y \longleftrightarrow (y - \pi R)$ -- and also
the strengths of the terms are not the special values chosen in
mUED.  The breaking of the $Z_2$ symmetry has far-reaching
consequences. For example, the $n = 1$ KK-modes of the neutral
gauge bosons, $B^{1}$ and $W_3^1$, may be produced singly at the
LHC and can decay to zero-mode fermion-antifermion pairs. We
explore the prospects of detecting a signal of KK-particles
through this route at the LHC.

\section{Boundary-localized terms, nmUED}

When boundary-localized terms (BLT) come
into play it is convenient to express four-component
5-dimensional fermion fields using two component chiral
spinors\footnote{The Dirac gamma matrices are in the chiral
representation with $\gamma_5 = diag( -I, I)$.}  as
\cite{schwinn}:
\begin{equation} 
\Psi_L(x,y) = \pmatrix{\phi_L(x,y) \cr \chi_L(x,y)} 
=   \sum^{\infty}_{n=0} \pmatrix{\phi_n(x) f_L^n(y) \cr \chi_n(x) g_L^n(y)}
\;\; , 
\label{fiveDL}
\end{equation} 
\begin{equation} 
\Psi_R(x,y) = \pmatrix{\phi_R(x,y) \cr \chi_R(x,y)} 
=   \sum^{\infty}_{n=0}  \pmatrix{\phi_n(x) f_R^n(y) \cr \chi_n(x) g_R^n(y)} 
\;\;  . 
\label{fiveDR}
\end{equation} 
Above, the $SU(2)_L$ behaviour of the fields are suppressed.  In
mUED $f^n_i(y), ~g^n_i(y), ~(i =L,R)$ are either a sine or a
cosine function of $y$. For nmUED this will no longer be the
case. Further, the mass of the KK-excitations will deviate from
the simple $n/R$ formula.

In UED the scalar and vector boson fields are obviously 
also 5-dimensional. Their KK expansions can be similarly written down.

In nmUED one may additionally consider kinetic and mass terms
localized at the fixed points of the orbifold \cite{Dvali,
carena, delAguila,  flacke, asesh}. We restrict
ourselves to boundary-localized kinetic terms (BLKT) only.
Specifically, we examine the interaction of quarks and leptons
with electroweak gauge bosons in a 5-dimensional theory with BLKT
at $y = 0$ and $y = \pi R$.

We consider fermion fields
$\Psi_{L,R}$ whose zero-modes are the chiral projections of the
SM fermions. The five-dimensional free
fermion action with BLKT is written
as \cite{schwinn}:
\begin{eqnarray} 
S\!\!\!\!\! &=& \!\!\!\!\!  \int d^4x ~dy \left[ \bar{\Psi}_L i
\Gamma^M \partial_M \Psi_L
+ r^a_f \delta(y) {\phi} ^\dagger _L i \bar \sigma^\mu
\partial_\mu \phi_L + \right. \nonumber \\
&&\!\!\!\!\!\!\!\!\!\!\!\!\!\!\!\!\!\! \left. r^b_f \delta(y - \pi R)
{\phi} ^\dagger _L i \bar \sigma^\mu
\partial_\mu \phi_L + (\Psi_L, \phi_L, \bar \sigma
\leftrightarrow \Psi_R, \chi_R, \sigma) \right] \nonumber \\ &&  
\end{eqnarray} 
with ${\sigma}^\mu
\equiv (I, \vec{\sigma})$ and $\bar{\sigma}^\mu
\equiv (I, -\vec{\sigma})$; $\vec{\sigma}$ being the 
Pauli matrices.  $r^a_f, r^b_f$ are the strengths
of the boundary terms choosen to be the same for $\Psi_L$
and $\Psi_R$. In the following only the $\Psi_L, \phi_L$ part of
the lagrangian is considered. The results for $\Psi_R, \chi_R$
are similar.

Variation of the above action leads to
\begin{eqnarray}
& \left[1 + r^a_f \delta(y) + r^b_f \delta(y - \pi R) \right] m_n f_L^n - 
\partial_y g_L^n = 0,\nonumber \\
& m_n g_L^n + \partial_y f_L^n = 0, \; (n = 0,1,2, \ldots).
\end{eqnarray}
Eliminating $g_L^n$  one obtains:
\begin{equation}
\partial_y^2 f_L^n + \left[1 + r^a_f \delta(y) + r^b_f \delta(y - \pi R) 
\right] m_n^2 f_L^n = 0.
\end{equation}
The boundary conditions at $y = 0$ we impose are \cite{carena}:
\begin{eqnarray}
f^n(y)|_{0^-}\!\!\!\!\!\!&=&\!\!\!\!\!\! f^n(y)|_{0^+},\;\
\frac{df^n}{dy}\bigg|_{0^+} -
\frac{df^n}{dy}\bigg|_{0^-} \!\!\!\!= \!\! -r_f^a m_n^2 f^n(y)|_{0},
\nonumber \\&&
\end{eqnarray}
and similar conditions at $y = \pi R$ with $r_f^a \rightarrow r_f^b$.
Here we consider the alternative\footnote{For
another alternative see \cite{ddrs1}.} $r_f^a \equiv r_f \neq 0,
r_f^b = 0$. The solution for $0 \leq y < \pi R$ is:
\begin{equation}
f^n(y) = N_n \left[ \cos (m_n y) - \frac{r_f^a m_n}{2} \sin (m_n
y) \right] \;,   
\label{sol1}
\end{equation}
where $m_n$ for  $n = 0,1, \ldots  $ 
satisfy the transcendental equation \cite{carena}:
\begin{equation} 
\tan(m_n \pi R)=-\frac{r^a_f m_n}{2} .
\label{trans}
\end{equation} 
The solutions
satisfy the {\em orthonormality} relations:
\begin{equation}
\int dy \left[1 + r^a_f \delta(y))
\right] ~f^n(y) ~f^m(y) = \delta^{n m}.
\end{equation}
The constant $N_n$ is determined from orthonormality:  
\begin{equation}
 N_n = \sqrt{\frac{2}{\pi R}}\left[ \frac{1}{\sqrt{1 + \frac{r_f^2 m_n^2}{4} 
+ \frac{r_f}{2 \pi R}}}\right].
\label{norm2}
\end{equation}

The reader is urged to note that (a) solutions in eq. (\ref{sol1}) 
are  combinations of sine and cosine
functions rather than any one of them alone and that (b) the
KK masses are solutions of  eq. (\ref{trans}) not simply
$n/R$. These features provide the novelty of nmUED.

In our work  we  only require the zero-modes and the $n = 1$
excitations of the fermion fields. We also deal with the
five-dimensional SM Higgs scalar  and the electroweak gauge
bosons.  For these boson fields one can undertake a similar KK
decomposition.  For the Higgs field we take $r_h^a \equiv r_h
\neq 0, r_h^b = 0$ while for the gauge fields $r_G^a \equiv r_G
\neq 0,  r_G^b = 0$. The upshot is that the $y$-dependent part of
the wave functions and the masses satisfy eqns. (\ref{sol1}) and
(\ref{trans}).

\begin{figure}[thb] 
{\vskip -1.5cm}
{\hskip -1.00cm}
\includegraphics[scale = 0.25, angle=0]{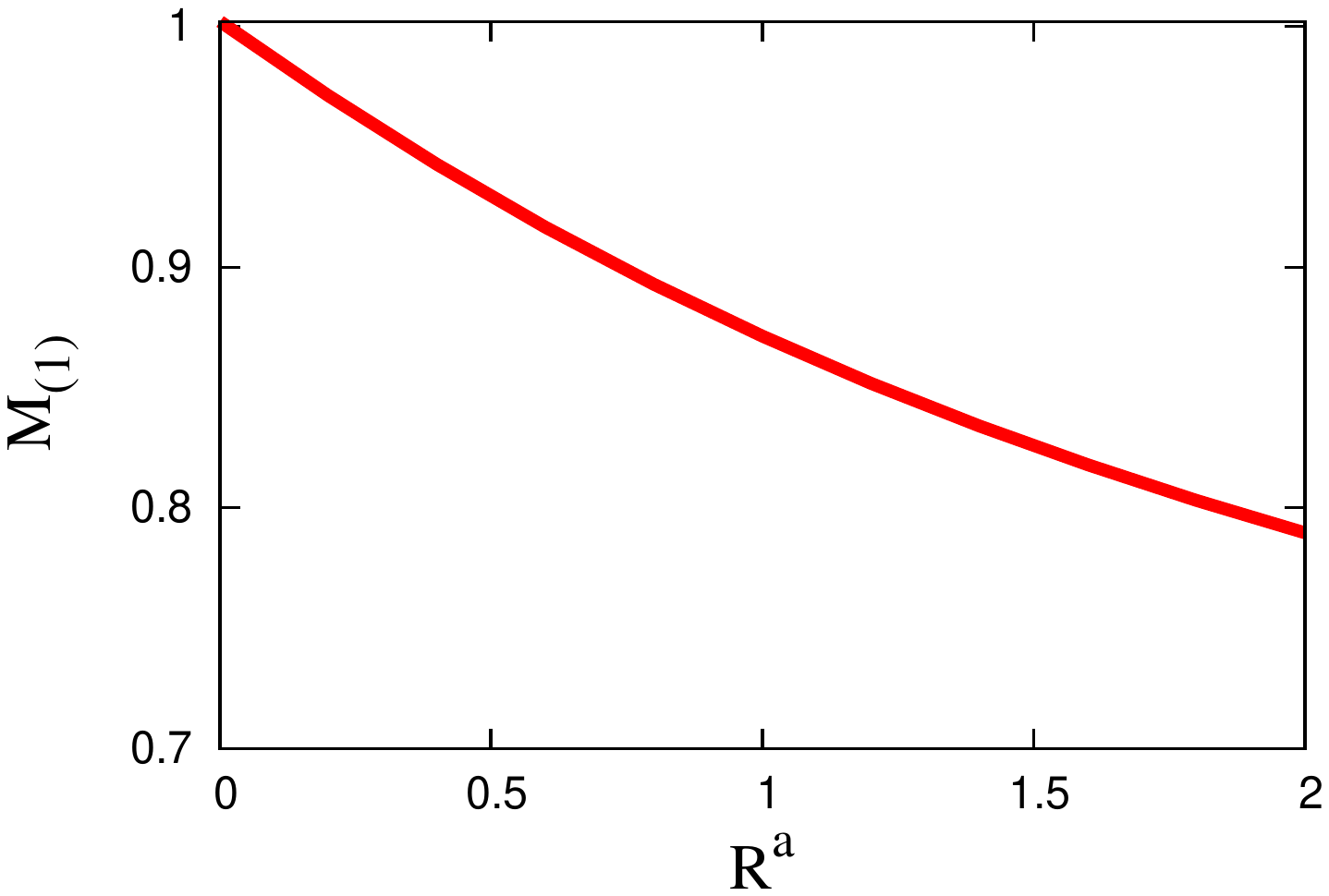}
{\vskip -13.7cm}
{\hskip 2.7cm}
\includegraphics[scale = 0.5, angle=0]{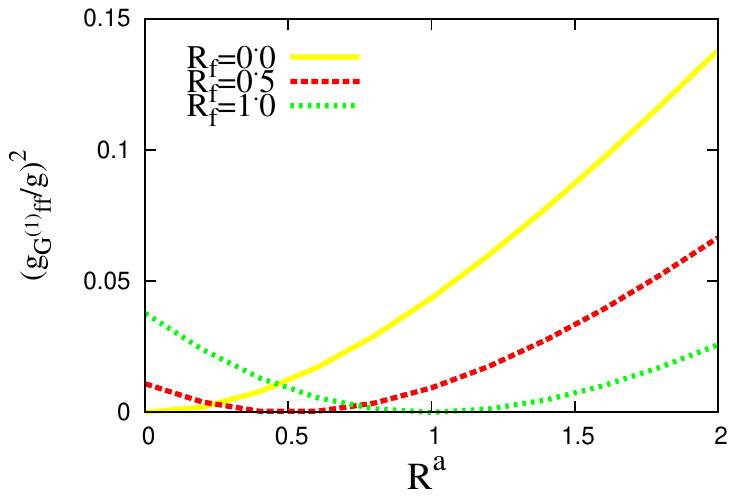}
\vskip -0.5cm
\caption{Left panel: Dependence of the mass of the $n =1$ KK-mode
$M_{(1)} \equiv m^{(1)} R$ on the BLKT strength $R^a \equiv r^a/R$
for any particle.  Note that larger $R^a$ yields a smaller mass.
Right panel: The square of the KK-parity violating coupling
(between $G^1 \equiv B^1 ~{\rm or} ~W_3^1$ and a pair of
fermions) with $R^a \equiv R_G$ for different choices of $R_f$.  }
\label{KKmass} 
\end{figure} 

The roots of equation (\ref{trans}), which may be obtained
numerically, are the extra-dimensional contributions to the
masses of the KK modes\footnote{The KK excitations also receive a
contribution to their masses from spontaneous breaking of the
electroweak symmetry.}.  In the left panel of Fig. \ref{KKmass},
we plot the dimensionless quantity $M_{(1)} \equiv m^{(1)} R$.
It applies to all fields, namely, scalars, fermions, and gauge
bosons. We show the variation of $M_{(1)}$ with\footnote{Here $R_f =
r_f/R$, $R_G = r_G/R$.} $R^a \equiv r^a/R$.  When $R^a = 0$, i.e.,
no BLKT at all, one gets $m^{(1)} = R^{-1}$, as expected.  One
finds that $m^{(1)}$ monotonically decreases as $R^a$  increases
asymptotically approaching 0.5$R^{-1}$.

It is to be noted that the mass $M_{(1)}$ is determined entirely
by the BLKT parametr, $R^a$, and the compactification radius
$R$. The gauge coupling is not involved. Therefore this
discussion applies for both $W_3^1$ and $B^1$ so long as the
appropriate BLKT parameters are used.

We have checked that for the range of BLKT which we
entertain the mixing between states of different KK level, $n$,
is very small and may be ignored. Further, only if the BLKT
parameters for the $B$ and $W$ gauge bosons are equal or very
nearly equal the mixing between $B^1$ and $W_3^1$ is substantial,
it being equal to the zero-mode weak mixing angle in the case of
equality. If $(r_B - r_W)/R$ is as small as 0.1 this mixing is
already negligible.  This follows from the mass matrix as we
outline below.

The mass matrix for the $n=1$ neutral electroweak gauge bosons
receives contributions from two sources: one originates from the
spontaneous breaking of the electroweak symmetry and the
other due to the  extra-dimension discussed above.
The contributions from symmetry breaking are ${\cal O}(v^2)$. The order of the
extra-dimensional contributions, ${(m_G^{(1)})}^2$, is set by
$(1/R)^2$ and is always much larger by far.  Thus effectively
these terms determine the mass eigenvalues
and the mixing  is negligible for $(R_W - R_B) \sim 0.1$ or
larger\footnote{If $R_W = R_B$ then the dominant diagonal terms
are equal and do not contribute to the mixing and simply shift
the masses of the eigenstates. In this case the mixing between
$W_3^1$ and $B^1$ is just as in the Standard Model with $\tan
\theta = g'/g$.}.  So,  we take $B^1$ and
$W_3^1$ to be the neutral electroweak gauge eigenstates for the
$n = 1$ KK-level.

The five-dimensional gauge couplings $g_5, g_5'$ and the
vacuum expectation value (vev) $v_5$ are related to the standard
couplings $g, g'$ respectively and the vev $v$ defined in four
dimensions by:
\begin{eqnarray}
g_5 \!\!\!\!\! &=& \!\!\!\!\! g \sqrt{\pi R ~S_W}  ,  ~g'_5 = g' \sqrt{\pi R ~S_B} , v_5 =
v/\sqrt{\pi R ~S_H} , \nonumber \\ &&   
\end{eqnarray}
where
\begin{eqnarray}
S_{W} \!\!\!\!\! &=& \!\!\!\!\! \left(1+\frac{R_W}{2\pi}\right),  
S_B = \left(1+\frac{R_B}{2\pi}\right),
S_H = \left(1+\frac{R_h}{\pi}\right). \nonumber \\ &&
\end{eqnarray}

\section{Coupling of $B^{1}$ and $W^{1}_3$ with zero-mode fermions}

We have now all the ingredients needed to calculate the coupling
of the states $W_3^{1}$ and $B^{1}$ to two zero-mode fermions
$f^{(0)}$.  Here $f^{(0)}$ could be SM quarks or leptons.  
We find that the results are not
very sensitive to the exact value of $R_h$. Our chosen value
ensures that the $H^{1}$ is always heavier than the $B^{1}$ and
$W_3^1$. 

Below we discuss the case for a generic gauge
boson $G^1$, which could be either of $B^1$ or $W_3^1$. The
$y$-dependent wave-functions of our interest here are found to be
\begin{equation} 
f_L^{0} = g_R^{0} = \frac{1}{\sqrt{\pi R(1 + R_f/2 \pi)}}, 
\end{equation}
for fermions and for the $n = 1$ gague boson
\begin{eqnarray}
a^{1} = \sqrt{\frac{1}{\pi R}}~\sqrt{\frac{2}{1+\left(\frac{R^a
M_{(1)}}{2}\right)^2+\frac{R^a}{2\pi}}} \nonumber \\
\left[\cos\left(\frac{M_{(1)}y}{R}\right)-
\frac{R^a M_{(1)}}{2}\sin\left(\frac{M_{(1)} y}{R}\right) 
\right] \;,
\end{eqnarray} 
where $R^a \equiv r^a/R$. Using the above we find 
\begin{eqnarray} 
g_{G^{1}f^{0}f^{0}}\!\!\!\!\! &=& \!\!\!\!\!    
\frac{\sqrt{2} ~g(G) \sqrt{S_G}}
{\left(1+\frac{R_{f}}{2\pi}\right) 
\sqrt{1+\left(\frac{R^a M_{(1)}}{2}\right)^2+\frac{R^a}{2\pi}}}  
\left(\frac{R_{f}-R^a}{2\pi}\right). \nonumber \\ &&
\label{coup2}
\end{eqnarray} 
where  we have used as earlier $M_{(1)} \equiv m_{G}^{(1)}  R$.

In the right panel of Fig. \ref{KKmass} we plot the square of the
above coupling strength as a function of $R^a$ for several
choices of $R_f$.  A noteworthy feature is that the coupling
vanishes when $R^a = R_f$.

\section{$B^1$ or $W_3^1$ production at the LHC and decay} 

We now turn to discuss a signal of nmUED at the LHC.  We are
interested in the resonant production of the $n=1$ KK-modes of
neutral EW gauge bosons, by the process $p p$ $(q \bar q)
\rightarrow G^{1}$ followed by $G^{1}
\rightarrow l^{+}l^{-}$ where $G^1$ is either of $B^1$ and $W_3 ^1$ 
and $l^\pm$ could be either $e^\pm$ or $\mu^\pm$. 

From here onwards for the SM particles we will not explicitly
write the KK-number $(n = 0)$ as a superscript.  The final state
leads to two leptons $(e ~{\rm or} ~\mu)$, with invariant mass
peaked at $m_{G^1}$. Note that the production as well as the
decay of these $n=1$ KK-excitations are driven by KK-parity
violating couplings.   If such a signature is found at the LHC,
then it would be a good channel for the determination of such
KK-parity violating couplings.

The production cross section in
proton proton collisions can be written in a compact form :
\begin{eqnarray}
\sigma (p p \rightarrow G^{1} + X) = \frac{4 \pi^2}{3 m_{G^1}^3}\;\sum_i 
\Gamma(G^{1} \rightarrow q_i \bar q_i) \nonumber \\
\int_\tau ^1 \frac{dx}{x}\;
\left[f_{\frac{q_{i}}{p}}(x,m_{G^1}^2) 
f_{\frac{{\bar q_{i}}}{p}}(\tau/x,m_{G^1}^2) + 
q_i \leftrightarrow \bar q_i \right]
\label{x-sections}
\end{eqnarray}

Here, $q_i$ and $\bar{q_i}$ represent a generic quark and
the corresponding antiquark of the $i$-th flavour respectively.
$\Gamma(G^{1} \rightarrow q_i \bar q_i)$ is the decay
width of $G^1$ into a quark and antiquark pair of the $i$th
flavour. $\tau \equiv {m_{G^1}^2/S_{PP}}$, where $\sqrt{S_{PP}}$
is the proton proton c.m. energy. $f$ stands for the  
quark or antiquark distribution functions within a proton.

For $B^1$ production, 
\begin{equation}
\Gamma = (g_{G^1 q
\bar{q}}'^2/32 \pi) \left[(Y^{q}_L)^2 + (Y^{q}_R)^2 \right]
m_{B^1}
\end{equation}
(with $Y^q _L$ and $Y^q _R$ the weak-hypercharges
for the left- and right-chiral quarks), while for producing $W^1 _3$ one
must use 
\begin{equation}
\Gamma = (g_{G^1 q \bar{q}}^2/128 \pi) m_{W^1_3}.
\end{equation}
$g^{(')}_{G^1 q \bar{q}}$ is the KK-parity violating coupling
among SM quarks and $W_3^1$ ($B^1$) as given in eq.
(\ref{coup2}). In the above 
equations $m_{G^1}$ stands for the mass eigenvalue of the gauge
boson $n=1$ excitation. 

The numerical results for the cross sections are obtained
using a parton-level Monte Carlo   with parton distribution functions
parametrized as in CTEQ6L \cite{CTEQ}.  We take the
$pp$ c.m. energy to be 8 TeV.  The renormalisation (for
$\alpha_s$) and factorisation scales (in the parton
distributions) are set at $m_{G^1}$.

To make a realistic estimate of the signal cross section, a simple 
calorimeter simulation has been implemented with:
\begin{itemize}
\item $p_T ^{\ell} >$ 20 ~GeV.
\item The calorimeter rapidity coverage  (for leptons) is $\vert
\eta \vert < 3.0$.

\item A cone algorithm with $\Delta R$ = $\sqrt {\Delta\eta^2 +
\Delta\phi^2}= 0.5 $ has been used for lepton isolation.
\end{itemize}

Once produced, $B^1$ ($W_3 ^1$) will decay {\em via} KK-parity
violating couplings to SM quarks and leptons and, if
kinematically allowed, to $f^1 \bar f$ (or
to $f \bar f^1$) through KK-conserving couplings. The
KK-parity violating leptonic decays provide a cleaner
environment at the LHC.  For simplicity we assume an universal
coefficient $r_f$ for the BLKTs involving all SM fermions. If
kinematically allowed (broadly when $R_f > R_G$), the KK-conserving
decay rates can be substantial and in such situations the
branching ratios (BRs) for KK-violating
decay rates are small. 
However, when the KK-conserving decays are kinematically
disallowed, decays to a SM fermion anti-fermion pair are the only
possible modes and hence the branching ratios become independent
of the input BLKT strengths.  Consequently, the decay rate of
$B^1$ to any fermion species is proportional to the sum of the
squares of the respective weak hypercharges $\left[(Y^f_L)^2 +
(Y^f_R)^2 \right]$ of the left- and right-chiral components. $W_3
^1$, on the other hand, decays democratically with branching
ratio of $1 \over 21$ to each type of left-handed zero-mode
fermions.  This immediately implies that the decay branching
ratio of $B^1$ ($W_3 ^1$) to $e^+ e^-$ or $\mu^+ \mu^-$ is
approximately $30\over 103$ $\left(2\over 21\right)$.

Final states with dileptons arise in
the SM mainly from resonant $Z$-production or Drell-Yan (DY).
The first  of these can be vetoed as in this
case the dilepton invariant mass peaks around $m_Z$. We find
that for 10 GeV bins around 700 (800) GeV the DY cross section
is  2.29 (1.56) $fb$. This  background though non-negligible
is such that for the $W_3^1$ and $B^1$ masses which we consider
$S/\sqrt{B} \geq 5$ can be reached for 20 $fb^{-1}$ integrated
luminosity.

\begin{figure}[tb]
{\vskip -1.5cm}
{\hskip -0.50cm}
\includegraphics[scale = 0.25, angle=0]{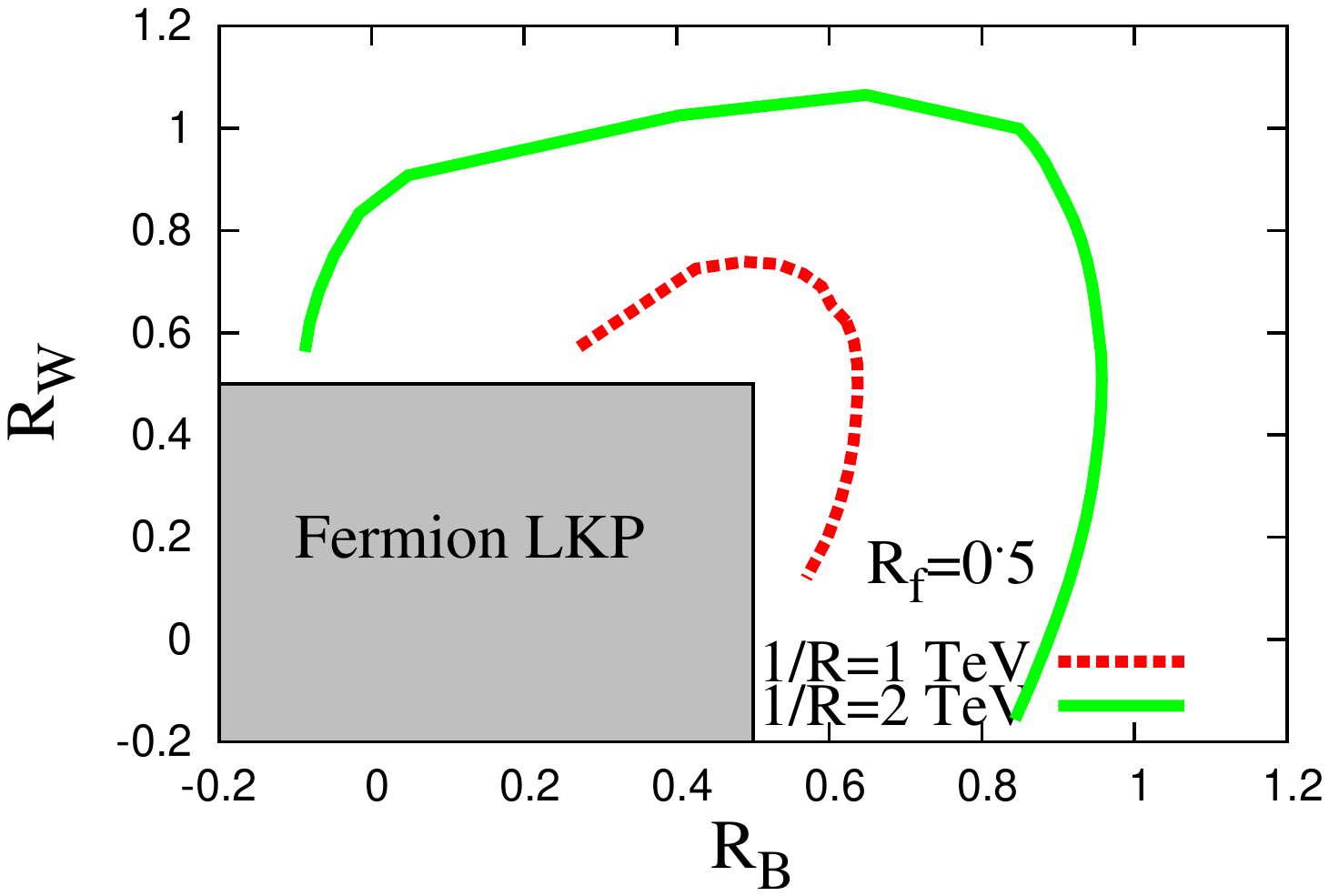} 
{\vskip -5.4cm}
{\hskip 3.5cm}
\includegraphics[scale = 0.25, angle=0]{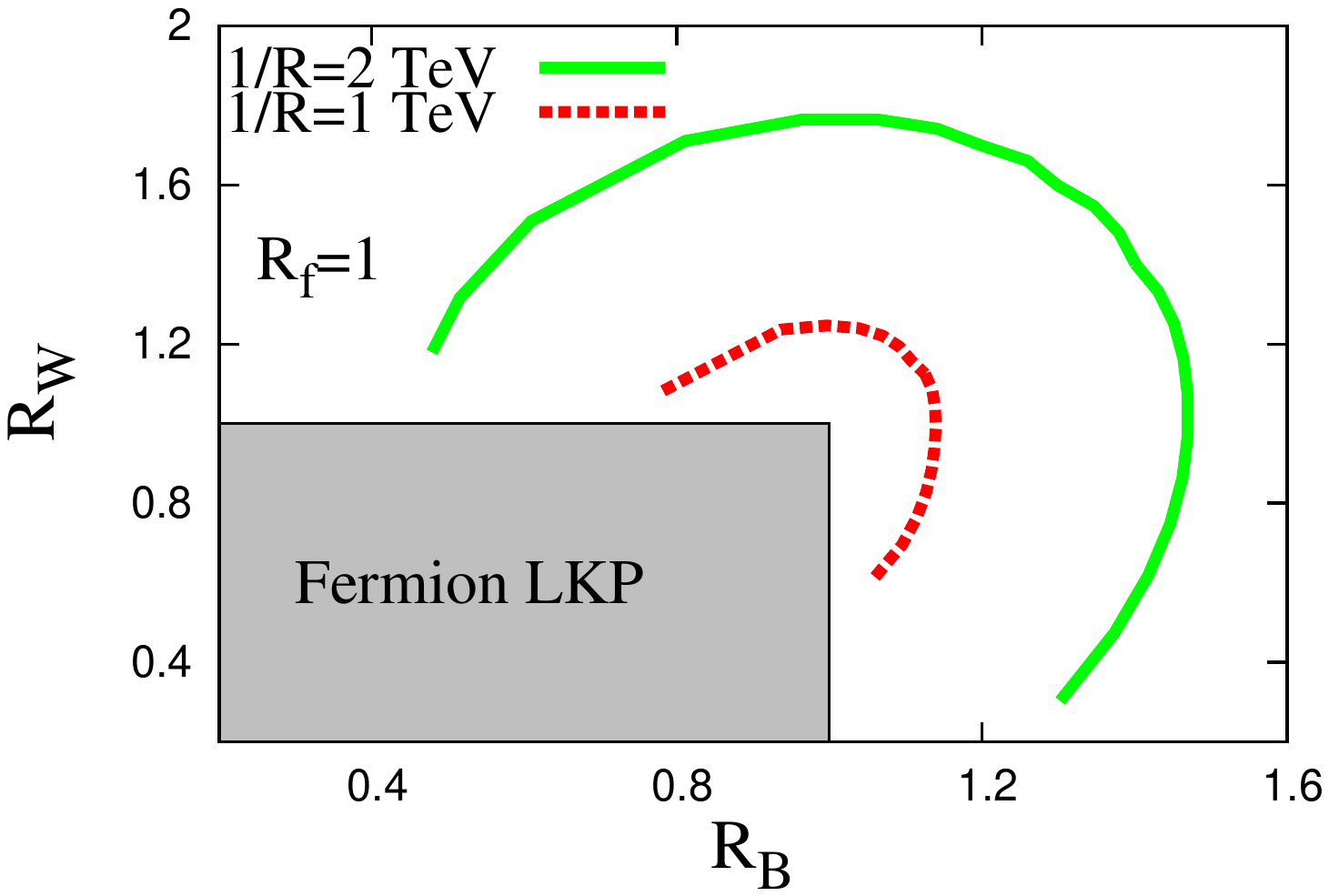}
\vskip -0.5cm
\caption{ Iso-event curves (40 signal events with 20 fb$^{-1}$
data at the LHC running at 8 TeV) for combined $W_3^1$ and $B^1$
signals in the $R_W - R_B $ plane. The region enclosed by the
solid green (dashed red) curve and the fermion LKP box yields
less than 40 events for $R^{-1} =$ 2 TeV (1 TeV). The left (right)
panel corresponds to  $R_f$ = 0.5 (1.0). Note difference in scales
in the two panels.}
\label{5fb_contoursS}
\end{figure}

We now present the results.  In Fig. \ref{5fb_contoursS} we have
plotted in the $R_B-R_W$ plane the iso-event curves\footnote{The
requirement here is that electron {\em plus} muon events together
resulting from $W_3 ^1$ {\em and} $B^1$ production and decay add
up to 40.} (40 events with 20 fb$^{-1}$ luminosity for LHC
running at 8 TeV).

The essence of Fig. \ref{5fb_contoursS} can be understood from
the earlier discussions. The KK-parity violating couplings vanish when $R_f =
R_G$ where $G \equiv W$ or $B$ as noted from  eq. (\ref{coup2}).
This is seen in the right panel of Fig.  \ref{KKmass} for
different choices of $R_f$. So, the production mode we consider
becomes unavailable. Further, in this situation, the $n$ = 1 gauge
boson and  fermion states are mass degenerate and KK-number
conserving decay modes are also not allowed.  In the
neighbourhood of this point there is an important asymmetry,
between whether $R_{B,W}$ is more than $R_f$ or less. In
the former case, the gauge boson $n$ = 1 state is lighter than the
corresponding fermion state and KK-number conserving decays are
not kinematically possible.  If both $R_{B}$ and $R_{W}$ are
smaller than $R_f$ then the fermion state is the LKP.

For  any iso-event contour the
largest value of $R_B$ will be for $R_W$ = $R_f$ and {\em
vice-versa}. This is because when $R_W$ = $R_f$ there is no
contribution to the signal from the $W^1_3$ as noted above.
As $R_W$ moves from this value the contribution from $W^1_3$
reduces the needed $R_B$.  This behaviour is not symmetric because of
the difference in the branching ratios of $W^1_3$ on the two
sides of $R_f$. 

In the figure we also indicate the region where the $n$ = 1
fermion is the LKP. It
is readily seen that for the point $R_W = R_B = R_f$ the
KK-number violating coupling is zero and also the $n=$1 states
are degenerate. So, no decays, neither KK-number violating nor
conserving, are permitted.

Data have been collected at LHC  at 8 TeV proton-proton c.m.
energy.  Observation or otherwise of the proposed  dilepton
signal would help in exploring or excluding the parameter space
of non-minimal UED models.  The projected exclusion limits can be
directly read off from the plots in Fig. \ref{5fb_contoursS}.
Any point above a particular iso-event curve can be excluded from
the non-observation of such events.    The $r_f$ dependence of
the cross section comes only through the coupling.  A careful
examination of the coupling in eqs.  (\ref{coup2}) shows that it
tends to a constant as $R_f
\rightarrow \infty$.

\section{Conclusions} 
To summarize, we have considered the effects of boundary
localized kinetic terms  in models where all  SM
fields can propagate in a spacetime with four spatial and one
timelike dimensions. The extra spatial dimension $y$ is compact
and can be considered as a circle of radius $R$ with a $y
\leftrightarrow -y$ symmetry. This results in two fixed
points at $y = 0$ and $y = \pi R$. At these points one
can include terms consistent with 4-dimensional Lorentz symmetry.
These are either kinetic  or mass terms. We examine
the former.

In the minimal Universal Extra Dimensional model, radiative
corrections play an important role in removing the
near-degeneracy of the masses of the KK-modes of  all SM
particles of the same KK-level, $n$.  UED, being an effective
theory, is valid only up to a cut-off, $\Lambda$.  In mUED the
boundary terms are fixed in a manner such that  at $\Lambda$ the
contribution due to radiative corrections is exactly compensated.
In fact, instead of calculating the radiative correction in a 5d
set up one may also parametrize these effects by incorporating a
set of BLKTs. These are somewhat similar in spirit to the
high-scale universal mass parameters $m_0$ and $m_{1\over2}$
often introduced in SUSY which serve as  boundary conditions for
the estimation of the low-energy masses.

There are two possibilities of choosing the BLKTs with rather
different physics consequences. In the first, the BLKTs are of
equal strength at the two boundary points $(y = 0, \pi R$).
Here, a $Z_2$ symmetry $y \longleftrightarrow (y - \pi R)$
remains.  This results in a theory where the spectrum of
KK-particles and the couplings are drastically different from
mUED. The lightest among the $n = 1$ KK particles can be a dark
matter candidate\footnote{An analysis of Dark Matter in the
context of these nmUED models is carried out in \cite{ddrs2}}.
The other alternative is to permit the BLKTs at $y=0$ and $y = \pi
R$ to be of unequal strengths.  This  leads to a breakdown of
KK-parity and will allow, for example, the decay we have
examined, $B^1 (W_3^1) \rightarrow e^+e^-, \mu^+\mu^-$, and
production of the $B^1 (W_3^1)$ singly.

In this presentation, we have examined the possible BLKTs for an
interacting theory of  fermions and the neutral
electroweak gauge bosons. We have considered the situation where
the BLKTs are present {\em only} at the $y = 0$ fixed point and
they are vanishing at $y = \pi R$. The boundary terms modify the
field equations for all particles in the $y$-direction.
Consistency conditions of the solutions of these equations
lead to masses of the higher KK-modes of fermions and the
photon.

For purpose of illustration, we have calculated the coupling
of $W_3^1$ and $B^1$, the $n = 1$ KK-excitations of the neutral
electroweak gauge bosons, to a
pair of zero-mode fermions (i.e.,  SM fermions) as a function of
$r_f, r_G, ~{\rm and} ~R^{-1}$.  The production and decay
of $W_3^1$ and $B^1$ at the LHC, {\em via} these KK-parity violating
couplings, have been considered. We have explored the
viability of the {\em dilepton} signature at the LHC
running at 8 TeV $pp$ center of mass energy.  It is
revealed that non-observation of such a high mass {\em dilepton}
signal  with 20 fb$^{-1}$ integrated luminosity in the 8 TeV run
of LHC will disfavour a large part of the parameter space
(spanned by $r_f, r_G, ~{\rm and} ~R^{-1}$).

The same signal can also arise if there are
extra $Z$-like bosons as in several popular extensions of the SM,
e.g., the Left-Right  symmetric models or models with an extra
$U(1)$ symmetry. We have not compared  the predictions of the
model under consideration with those in these other scenarios.

{\bf Acknowledgements} AD acknowledges partial support from the
UGC DRS project sanctioned to the Department of Physics,
University of Calcutta. UKD is supported by  funding from the
Department of Atomic Energy, Government of India for the Regional
Centre for Accelerator-based Particle Physics, Harish-Chandra
Research Institute.   AS is the recipient of a Junior
Research Fellowship from the University Grants Commission. AR is
thankful to the Department of Science and Technology for a J.C.
Bose Fellowship.




\nocite{*}
\bibliographystyle{elsarticle-num}
\bibliography{martin}




\end{document}